\documentstyle[12pt,psfig]{article}
\begin{document}
\begin{center}
{\Large \bf Study of the $\eta$-nucleus interaction in the
$p\, d
\rightarrow \, ^3$}{\Large He}{\Large \bf$ \,\eta$ reaction near
threshold}
\end{center}
\begin{center}
K. P. Khemchandani$^{}$\footnote{email: kanchanp@magnum.barc.ernet.in}
, N. G. Kelkar$^{}$\footnote{email: ngkelkar@apsara.barc.ernet.in}
 and B. K. Jain$^{}$\footnote{email: brajesh$_-$jain@vsnl.com}
\end{center}
\begin{center}
Nuclear Physics Division, Bhabha Atomic Research Centre,\\
Mumbai 400085,
India \\
\end{center}
\vskip0.2cm
\begin{abstract}
A detailed study of the effect of the $\eta$-nucleus final state
interaction
(FSI) in the $p\, d \rightarrow \, ^3$He$\, \eta$ reaction
close to threshold is
presented. The FSI is incorporated through a T-matrix for $\eta\,^3$He
elastic scattering, constructed using few body equations. This T-matrix
accounts for off-shell and binding effects in $\eta$-nucleus scattering.
The
energy dependence of the data on the $p\, d \rightarrow \, ^3$He$\, \eta$
reaction
near threshold is reproduced only after including the FSI. The off-shell
and binding
effects in $\eta\,^3$He scattering are found to be important. Given the
uncertainty in the knowledge of the elementary $\eta$-nucleon
interaction,
the sensitivity of the $ p\,d \rightarrow \, ^3$He$\,\eta$ cross section
to different prescriptions of the elementary t-matrix, $t_{\eta \, N \,
\rightarrow \, \eta \, N}$ is also discussed.
\end{abstract}
\noindent
PACs numbers: 13.75.-n, 25.40.Ve, 25.10.+s

\section{Introduction}
Experimental data on the $p\, d \rightarrow \, ^3$He$\, \eta$ reaction
close to
threshold have revealed some surprising features \cite{mayer,berger}.
Inspite of the large momentum transfer involved in $\eta$ production as
compared to that in
pion production, the cross section for $p\, d \rightarrow \, ^3$He$\,
\eta$ is
large and comparable with that for $p\, d \rightarrow \,^3$He$\, \pi^0$.
The
energy dependence of the two reactions near threshold is also very
different. The $p\, d \rightarrow \, ^3$He$\, \eta$ reaction shows a much
rapid
variation, with the threshold amplitude falling by a factor of 3.75 over
an
$\eta$ centre of mass momentum of 75 MeV/c.

The observed features of the $p\, d \rightarrow \, ^3$He$\, \eta$
reaction have
been attributed to the strong final state $\eta$-nucleus interaction.
This
reaction was studied in Ref. \cite{wilkin}, where the final state
interaction
(FSI) was incorporated in an approximate way through an enhancement
factor.
The cross section was factorized in terms of the amplitude for the
reaction
$p\, d \rightarrow \, ^3$He$\, \eta$ with plane waves
for the $\eta\,^3$He in the
final state and an S-wave FSI factor written in terms of the on-shell
$\eta$-nucleon amplitude. The energy dependence of the cross section data
was reproduced, though its predicted absolute value was lower by a factor
of
2.5. In yet another work \cite{wycech}, the scattering length of the
$\eta$
meson on helium was calculated using multiple scattering theory, which
was
then used to calculate the FSI factor for the
$p\, d \rightarrow \, ^3$He$\, \eta$
cross section. The FSI factor with an $\eta$-nucleon scattering length
$a_{\eta N} = (0.291, 0.36)$ leading to $a_{\eta\,^3He}$=(-0.89, 1.8)
was found to give a good fit to the $p\, d \rightarrow \, ^3$He$\, \eta$
data.

The $p\, d \rightarrow \, ^3$He$\, \eta$ reaction
at threshold and higher energies
has also been studied theoretically to investigate the reaction mechanism
involved \cite{laget,kondrat,bkj}. Due to the large mass of the $\eta$
meson,
the momentum transfer involved in this reaction is large.
It is 900 MeV/c at threshold and reduces to 500 MeV/c by about 1 GeV above 
threshold.
As a result of this, it was shown in Ref. \cite{laget}
that the three-body mechanism which allows the momentum transfer to be
shared amongst three nucleons dominates. The one and two body mechanisms
were found to underestimate the experimental cross sections by more
than two orders of magnitude.
The three nucleons share the large momentum transfer through a
two step process where the incident proton interacts with a nucleon in
the
deuteron to produce a pion which then interacts with the other nucleon in
the deuteron to produce an $\eta$ meson. The
$p\, d \rightarrow \, ^3$He$\, \eta$
reaction, thus,  proceeds via the $N N \rightarrow \pi d$ and $\pi N
\rightarrow \eta N$ reactions. 
Though the $p\, d \rightarrow \, ^3$He$\, \eta$
reaction can in principle proceed through the one
and two body mechanism, considering the support in existing literature
\cite{laget,kilian} for the three body mechanism and the fact that the low
momentum components of the nuclear wave function are picked up in this way,
at least at energies very close to threshold it seems justified to consider
this mechanism to be the only major reaction mechanism for the
$p\, d \rightarrow \, ^3$He$\, \eta$ reaction.

In the present work we study the $p\, d \rightarrow \, ^3$He$\, \eta$
reaction near threshold using the three body mechanism mentioned above.
Our main objective is to investigate
the FSI in this reaction
in a rigorous way. The above mentioned theoretical works
in the literature either neglect the
FSI or incorporate it in an approximate way using on-shell amplitudes. We
express the $\eta\,^3$He relative wave function in terms of the Lippmann
Schwinger equation involving the T-matrix for $\eta\,^3$He elastic
scattering.
This T-matrix is evaluated using a method of few body equations
\cite{bela,rakit1,rakit2} which will be described in
detail in the next section.

In section 2 we present the formalism for the calculation of the
$p\, d \rightarrow \, ^3$He$\, \eta$ cross section including the FSI.
The production mechanism for $p\, d \rightarrow \, ^3$He$\, \eta$
reaction
within a two-step model is described in section 3. In section 4 we
present the results. We reproduce the shape and magnitude of the
experimentally measured scattering amplitude. The off-shell effects in
$\eta\,^3$He scattering are found to be important in producing the energy
dependence of the cross section.

\section{Final state interaction}
The transition matrix for the reaction $p\, d \rightarrow \, ^3$He$\,
\eta$,
which includes the interaction between the $\eta$ meson and $^3$He is
given
by,
\begin{equation}\label{tmat}
T = \, \,  <\,\Psi_{\eta \,^3He}^-\,(\vec{k_\eta})\, ; \, m_3\, | \, T_{p
d \rightarrow
\,^3He \,\eta}\,| \, \vec{k_p}\, ; \, m_1 \, m_2 \,>
\end{equation}
where $m_1$, $m_2$ and $m_3$ are the spin projections of the proton,
deuteron
and helium respectively. $ \vec {k_p}$ and $ \vec {k_\eta}$ are the
momenta
of the particles in the initial and final states. The final state
$\eta\,^3$He wave function $\Psi_{\eta\,^3He}^{-*}$ consists of a plane
wave
and a scattered wave, and can be written as,
\begin{equation}\label{wave}
<\,\Psi_{\eta \,^3He}^- \,| \, =\, <\, \vec{k_{\eta}}\,| \, + \,
\int { d\vec {q} \over (2 \pi)^3 } \, {<\,\vec{k_{\eta}} \,|\,
T_{\eta \,^3He}\, | \, \, \vec{q}\,>
\over E(k_{\eta}) \, - E(q)\, + \,i\epsilon} \, <\vec{q}\,|
\end{equation}
where $T_{\eta\,^3He}$ is the T-matrix for $\eta\,^3$He elastic
scattering.
Replacing the above wave function in Eq. (\ref{tmat}), we get,
\begin{eqnarray}\label{tmat2}
&&T =\, <\,\vec{k_\eta}\, ; \, m_3\,|\, T_{p d \rightarrow \,^3He
\,\eta}\,|
\,\vec{k_p}\, ; \, m_1 \, m_2\,> + \\ \nonumber
&&\sum_{m_3^\prime} \int { d\vec {q} \over (2\pi)^3} {<\, \vec{k_\eta}\,
; \, m_3\,|
\, T_{\eta\, ^3He}\, |\,\, \vec{q}\, ; \, m_3^\prime\,> \over E(k_\eta)\,
- \,E(q)\,
+\, i\epsilon}\,
<\vec{q}\, ; \, m_3^\prime\,| T_{p d \rightarrow \,^3He \,\eta}\,|
\,\vec{k_p}\, ; \, m_1 \, m_2>
\end{eqnarray}
The matrix elements $< |T_{p \,d \rightarrow \,^3He\, \eta}| >$ in the
above equation correspond to the Born amplitude for the
$p\, d \rightarrow \, ^3$He$\, \eta$ reaction.
We calculate these matrix elements
using a two step model which will be discussed in the next section.

The T-matrix, $T_{\eta\,^3He}$,
in Eq. (\ref{tmat2}) for $\eta\,^3$He elastic scattering is
evaluated using four particle equations for
the $\eta (3N)$ system. For practical convenience, the evaluation of
$T_{\eta\,^3He}$ is done within a Finite Rank Approximation (FRA)
approach.
This means that the nucleus in the elastic meson-nucleus
scattering is always in its ground state.
The shortcomings of the FRA for the $\eta$-deuteron system have been
investigated in Ref. \cite{shevchenko}. However, for the low energies concerned
in this work and also as mentioned in Ref. \cite{shevchenko},
it seems justified to use the FRA for the
$\eta-\,^3$He system. We write the target Hamiltonian $H_A$ as,
\begin{equation}\label{hamilt}
H_A \approx \varepsilon |\psi_0> <\psi_0|
\end{equation}
where $\psi_0$ is the nuclear ground state wave function and
$\varepsilon$
the binding energy.
 
Within this approximation, the $\eta\,^3$He T-matrix is
given as \cite{bela,rakit1,rakit2},
\begin{eqnarray}\label{tfsi}
T(\vec{k^\prime},\, \vec{k}\,; z) &=& <\, \vec {k^\prime}\, ; \,
\psi_0\,|\, T^0(z)
\, | \, \vec{k} \, ; \, \psi_0\,> \, +\, \\ \nonumber
&&\varepsilon\, \int {\vec{dk^{\prime\prime}}
\over (2\pi)^3} {<\,\vec{k^\prime}\, ; \,  \psi_0 \,|\, T^0(z)\, | \,
\vec{k^{\prime\prime}}\,
; \,  \psi_0\,> \over (z - {k^{\prime\prime\,2} \over 2\mu})(z -
\varepsilon
- {k^{\prime\prime\,2} \over 2\mu})}
T(\vec{k^{\prime\prime}},\, \vec{k}\, ; \, z)
\end{eqnarray}
where $z = E - |\varepsilon| + i0$. $E$ is the energy associated with
$\eta$-nucleus relative motion and $\mu$ is the reduced mass of the
$\eta$-nucleus system. The operator $T^0$ describes the
scattering of $\eta$ meson from nucleons fixed in their space position
within the nucleus. The matrix elements for $T^0$ are given as,
\begin{equation}\label{t0mat}
<\, \vec{k^\prime} \, ; \,  \psi_0\,|\, T^0(z)\, |\, \vec {k} \, ; \,
\psi_0\,> =
\int d\vec{r}\, |\, \psi_0(\vec{r})\, |^2 \, T^0\, (\vec{k^\prime},\,
\vec{k}\,;
\vec{r}\,;z)
\end{equation}
where,
\begin{equation}\label{t0mat2}
T^0\,(\vec{k^\prime},\, \vec{k} \,;\vec{r} \,;z) = \sum_{i=1}^A \,
T_i^0\,
(\vec{k^\prime},\, \vec{k}\,;\vec{r_i}\,;z)
\end{equation}
$T_i^0$ is the t-matrix for the scattering of the $\eta$-meson from the
$i^{th}$ nucleon in the nucleus, with the rescattering
from the other (A-1) nucleons included. It is given as,
\begin{equation}\label{t0mat3}
T_i^0\,(\vec{k^\prime},\, \vec{k}\,;\vec{r_i}\,;z) =
t_i(\vec{k^\prime},\, \vec{k}\,;\vec{r_i}\,;z) + \int
{d\vec{k^{\prime\prime}}
\over (2\pi)^3}\,{t_i(\vec{k^\prime},\,
\vec{k^{\prime\prime}}\,;\vec{r_i}\,;z)
\over z - {k^{\prime\prime\,2} \over 2\mu}} \sum_{j\neq i}
T_j^0(\vec{k^{\prime\prime}},\, \vec{k}\,;\vec{r_j}\,;z)
\end{equation}
The t-matrix for elementary $\eta$-nucleon scattering, $t_i$, is written
in terms of the two body $\eta N$ matrix $t_{\eta N}$ as,
\begin{equation}\label{tetan}
t_i(\vec{k^\prime},\, \vec{k}\,;\vec{r_i}\,;z) =
t_{\eta\,N}(\vec{k^\prime},\,
\vec{k}\,;z)\, exp [\,i (\, \vec{k} -
\vec{k^\prime}\,)\cdot\,\vec{r_i}\,]
\end{equation}

The $^3$He wave function $\psi_0$, required in the
calculation of $T_{\eta\,^3He}$ is taken to be of the Gaussian form.

Since there exists a lot of uncertainty in the knowledge of the
$\eta$-nucleon interaction, we use three different prescriptions of
the $\eta$-N t-matrix, t$_{\eta \, N \, \rightarrow \, \eta \, N}$, 
leading to different values of the $\eta$-N scattering length. We give
a brief description of these models of 
t$_{\eta \, N \, \rightarrow \, \eta \, N}$ below. The elementary 
matrix, t$_{\eta \, N \, \rightarrow \, \eta \, N}$ is then used in
the evaluation of the T-matrix for $\eta\,^3$He scattering. In 
Ref. \cite{fix} a coupled channel t-matrix including the $\pi$N and 
$\eta$N channels with the S$_{11}$ - $\eta$N interaction playing a 
dominant role was constructed. The t-matrix thus consisted of the 
meson - N* vertices and the N* propagator as given below:
\begin{equation}\label{tfix}
t_{\eta \, N \, \rightarrow \, \eta \, N} (\, k^\prime, \, k; z) = 
{ { \rm g}_{_{N^*}}\beta^2 \over (k^{\prime\,2} +
\beta^2)}\,\tau_{_{N^*}}(z)\,{ {\rm g}_{_{N^*}}\beta^2 \over (k^2 + \beta^2)}
\end{equation}
with,
\begin{equation}\label{tau}
\tau_{_{N^*}}(z) = ( \, z - M_0- \Sigma_\pi(z) - 
\Sigma_\eta(z) + i\epsilon \, )^{-1}
\end{equation}
where $\Sigma_\alpha(z)$ $(\alpha = \pi, \eta)$ are the self energy
contributions from $\pi N$ and  $\eta N$ loops. The parameters of this model
were chosen to produce an $\eta$N scattering length, $a_{\eta N}$ = (0.75, 0.27) fm
which is in agreement with some other analyses available in literature 
\cite{green,batinic}. We will refer to this parameter set as `Fix {\it et al.} (I)' while
discussing the results.

In Ref. \cite{fix} the authors also perform calculations
with the choice of meson-nucleon cut-off parameters as in Ref. \cite{pena}. 
These parameters lead to $a_{\eta N}$ = (0.88, 0.41) fm. We refer to this parameter set 
as `Fix {\it et al.} (II)'.
We present our calculations for the $p \, d \, \rightarrow \,^3$He$\, \eta$ reaction
with the t-matrix of Ref. \cite{fix} using the two different sets of paramteres mentioned
above.

We also present results using the $\eta$-N t-matrix of Ref. \cite{bhal} which gives
a much smaller value of $a_{\eta N}$, namely, $a_{\eta N}$ = (0.28, 0.19) fm. In this model 
the $\pi$N, $\eta$N and $\pi \Delta$ channels were treated in a coupled channels formalism.
It is important to note that the parameters in all the coupled channel t-matrices
mentioned above are adjusted to reproduce the data on the 
$\pi \, N \, \rightarrow \, \eta \, N$ reaction, but the values of $a_{\eta N}$ deduced
from them are different.

\section{Production mechanism}
As mentioned in the Introduction, we assume the $\eta$ production
in $p\, d \rightarrow \, ^3$He$\, \eta$ to proceed through a two-step
process via the $N N \rightarrow \pi d$ and $\pi N \rightarrow \eta N$
reactions as shown in Fig. 1. The amplitude for the
$p\, d \rightarrow \, ^3$He$\, \eta$ reaction appearing in Eq.
(\ref{tmat2}) can be
written within this model as,
\begin{eqnarray}\label{ampli}
< |T_{pd \rightarrow ^3He\,\eta}| >=i \int {d\vec{p_1}\over (2\pi)^3}
{d\vec{p_2}\over (2\pi)^3} \sum_{int\,m's} <p n \,|\,d>
\,<\pi^+\,d |T_{pp\, \rightarrow\, \pi^+\, d}| p\,p>
\\ \nonumber
 \times{1\over (k_\pi^2-m_\pi^2+i\epsilon)}
\, <\eta\,p \,|\,T_{\pi N \rightarrow \eta p}\,| \pi^+\,n>
\,\,<\,^3He\,|\,p\,d>
\end{eqnarray}
\begin{figure} [h]
\centerline{\vbox{
\psfig{file=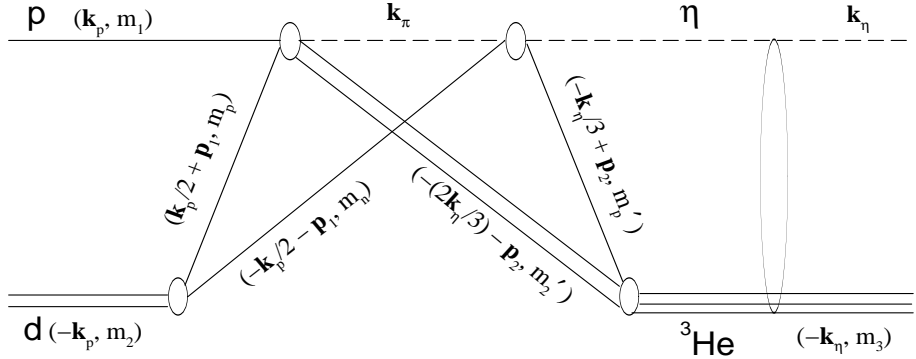,height=5cm,width=12cm}}}
\caption{Diagram of $\eta$ production in the
$p\, d \rightarrow \, ^3$He$\, \eta$
reaction with a two step process. The ellipse
indicates the final state interaction of $^3$He and $\eta$.}
\end{figure}
where the sum runs over the spin projections of the intermediate
particles. The spin projections and momenta of the interacting
particles are as shown in Fig. 1.
$k_{\pi}$ is the four momentum
of the intermediate pion which could either be \, $\pi^+$ or
$\pi^0$. In the case of an intermediate $\pi^0$, the matrix element
for $p\, d \rightarrow \, ^3$He$\, \eta$ is half of that written above
for
$\pi^+$. Hence we calculate the T-matrix as in Eq. (\ref{ampli})
and multiply it by a factor of 3/2 to account for the intermediate
$\pi^0$. Each of the individual matrix elements in the above equation
is expressed in terms of partial wave expansions. The matrix elements
for the $p p \rightarrow \pi^+ d$ reaction, parametrized in terms of
the available experimental data are taken from Ref. \cite{arndt}. For
the $\pi^+ n \rightarrow \eta p$ reaction, we use the coupled
channel t-matrix of Ref. \cite{bhal}, mentioned in the previous section.
The matrix elements $<p n|d>$ and $<^3$He$|p d>$ consist of the deuteron
and helium wave functions in momentum space. We use the deuteron
wave function from Ref. \cite{paris} where an analytical parametrization
of it was done with a Paris potential.
This wave function reproduces
the known low energy properties and the electromagnetic form factor
of the deuteron well. For the $^3$He wave function, we use the
parametrization given in Ref. \cite{helium}. The values of the
parameters in \cite{helium} were obtained by fitting the wave function
to the variational calculations of Schiavilla {\it et al.} \cite{schiav}
using the Urbana force. The details of Eq. (\ref{ampli}) are given
in the appendix.
 
\section{Results and Discussion}
\begin{figure}[h]
\centerline{\vbox{
\psfig{file=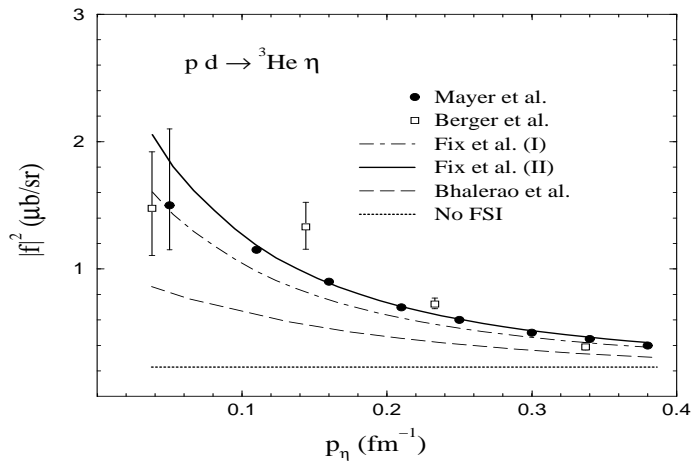,height=6.7cm,width=9cm}}}
\caption{The square of the $p\, d \rightarrow \, ^3$He$\, \eta$ amplitude
defined in Eq. (\protect\ref{fsquare}) as a function of the $\eta$
momentum in the centre of mass. The data is from Refs \protect\cite{
mayer,berger}. The dotted line is the calculation of the present
work without including the FSI. The solid, dash-dotted and dashed 
lines are the calculations including FSI with different prescriptions
of elementary t-matrices.}
\end{figure}
The reaction $p\, d \rightarrow \, ^3$He$\, \eta$ has been studied at
Saturne \cite{mayer,berger} for proton energies between 0.2 and 11 MeV above
threshold. Taking out the phase space factor, the spin averaged
amplitude can be defined as,
\begin{equation}\label{fsquare}
|\,f\,|^2 = {k_p \over k_\eta} \cdot {d\sigma \over d\Omega}_{cm}
\end{equation}
where $k_p$ and $k_{\eta}$ are the proton and $\eta$ momenta in the
centre of mass system.
The data on $|f|^2$ (see Fig. 2) drops rapidly
(by about a factor of 3.75 ) from threshold to 0.4 $fm^{-1}$
momentum (corresponding to 11 MeV energy) above
threshold. In Fig. 2 we compare our calculations of $|f|^2$
(at $\theta_{\eta} = 180^0$) with and
without the inclusion of the $\eta\,^3$He final state interaction (FSI),
with the data from Refs \cite{mayer,berger}. The dotted line
in Fig. 2 is our calculation without FSI and can be seen to be a constant
as a function of energy.
The dash-dotted, solid and dashed curves are the results obtained 
using the t-matrix of Ref. \cite{fix}, with parameter sets (I) and (II),
and that of Ref. \cite{bhal} respectively.
 We see that the FSI is
responsible for changing the shape of $|f|^2$ from a constant to
a rapidly falling one as a function of energy.

In what follows, we shall study the off shell and the binding energy effects
in the FSI. We also calculate the angular distribution and total cross-sections
for the $ p\,d\, \rightarrow \, ^3$He$ \, \eta$ reaction. In all these calculations
we use the prescription of Ref. \cite{fix} with parameter set (II) for the elementary t-matrix, 
$t_{\eta \, N\, \rightarrow \, \eta \, N}$. We arrive at the same results qualitatively,
in all the calculations if we choose the parameter set (I) of the same t-matrix or the 
elementary matrix, $t_{\eta \, N\, \rightarrow \, \eta \, N}$ of
Ref. \cite{bhal}.

\begin{figure}[h]
\centerline{\vbox{
\psfig{file=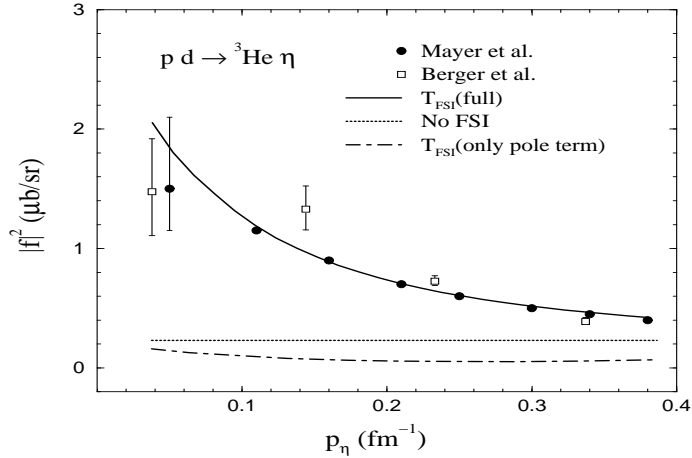,height=6.7cm,width=9cm}}}
\caption{The solid and dotted line and the data are as in Fig. 2. The dash
dotted line is the FSI calculation retaining only the pole term in
Eq. (\protect\ref{tmat2}). We have used the elementary t-matrix
of Ref. \cite{fix} with parameter set (II).}
\end{figure}
\begin{figure}[h]
\centerline{\vbox{
\psfig{file=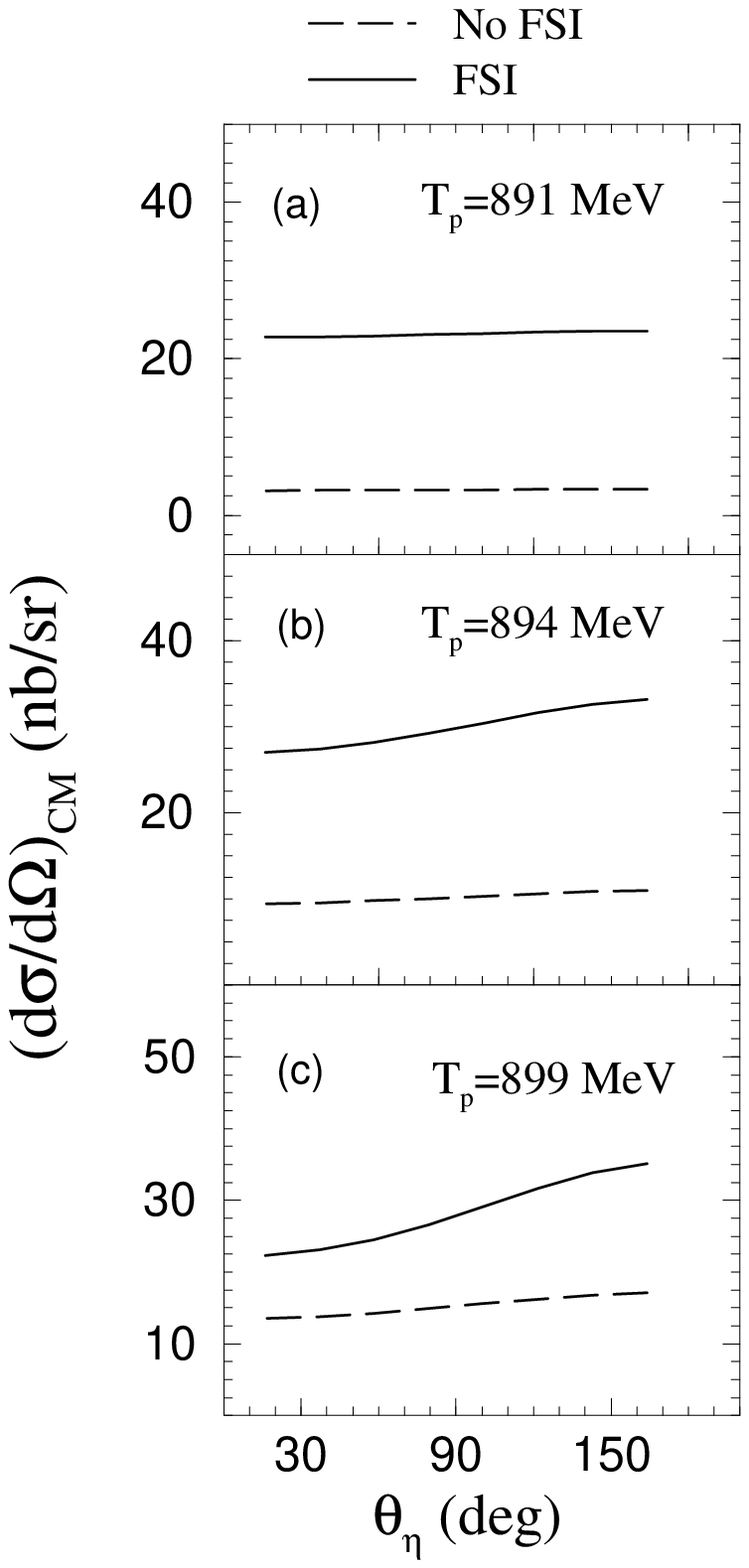,height=8.7cm,width=5cm}}}
\caption{Angular distribution of $\eta$ in the
$p\, d \rightarrow \, ^3$He$\, \eta$ reaction at different beam energies.
The
solid (dashed) curves are the calculations with (without) FSI included.}
\end{figure}
In Fig. 3 we study the off-shell effects in the FSI. As seen in
section 2, we describe the $\eta\,^3$He final state interaction through
an off-shell T-matrix for $\eta\,^3$He scattering. Previous estimates
of FSI in literature have been made using on-shell amplitudes and hence
it is important to check the validity of such an approximation.
The $\eta\,^3$He T-matrix appears in an integral in Eq. (\ref{tmat2})
which can be split into the principal value and pole term. Retaining
only the pole term in Eq. (\ref{tmat2}) and setting the principal
value (which involves the off-shell $\eta\,^3$He scattering) to zero,
we get the dash dotted line in Fig. 3. The pole term alone is unable
to reproduce the shape of the data and in fact reduces the magnitude
of the results obtained without FSI.
We find that 
 $t_{\eta \, N \, \rightarrow \, \eta \, N}$ of Ref. \cite{fix},
 with parameter set (I) and (II) and that of Ref. \cite{bhal} have 
similar on-shell behaviour meaning that the same dash dotted curve
of Fig. 3 represents the pole term calculation for the 3 different
prescriptions of $t_{\eta \, N \, \rightarrow \, \eta\, N}$. 
The solid line represents the full calculation using the prescription
of Ref. \cite{fix} with parameter set (II)
for the elementary t-matrix. Thus we see that including the off shell
effects produces the proper energy dependence of $|f|^2$ in addition
to increasing its magnitude as compared to the pole term calculation. 
The other two prescriptions of $t_{\eta \, N \, \rightarrow \, \eta \, N}$  
(Fix {\it et al.} (I) and Bhalerao {\it et al.}) show similar off-shell
effects. However, as can be seen from the curves in Fig. 2, 
the increase in magnitude as compared to the pole term is smaller. 
 
\begin{figure}[h]
\centerline{\vbox{
\psfig{file=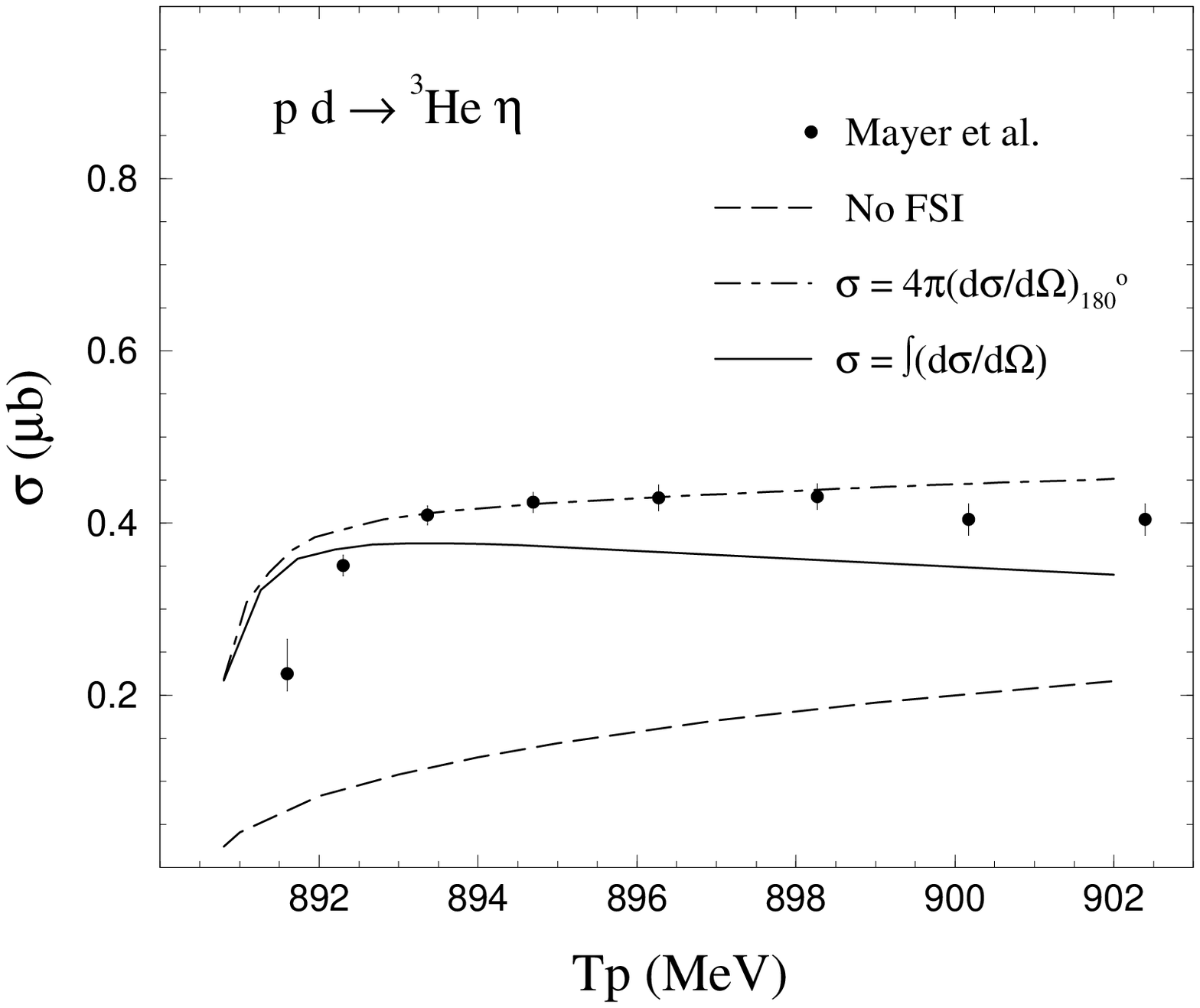,height=8cm,width=9cm}}}
\caption{Total cross section for $p\, d \rightarrow \, ^3$He$\, \eta$ as
a
function of beam energy. Solid curve is obtained by numerically
integrating
the calculated $d\sigma/d\Omega$ with FSI over all angles. The dash
 dotted curve
is obtained by multiplying $d\sigma/d\Omega$ (with FSI) at
180$^0$ with 4$\pi$, thus assuming the angular distribution to be
isotropic. The dashed curve is the calculation without FSI.
The data is from Ref. \protect\cite{mayer}.}
\end{figure}

Next, in Fig. 4, we study the effect of the FSI on the angular
distributions at different energies. 
As observed in Fig. 2 too, we see that the FSI increases the magnitude
of the cross sections with the increase being maximum at threshold
(Fig. 4a). The angular dependence of the cross sections without
FSI (dashed lines) is isotropic at threshold and deviates by a small
amount from the isotropy with increase in beam energy.
The small anisotropy at a few MeV above threshold is somewhat amplified
by the final state interaction. Thus,
as one goes to higher energies (Fig. 4c),
though the FSI does not increase the magnitude of the
cross section too much, it does change the angular dependence.

In Fig. 5 we show the angle integrated total cross section as a function
of beam energy. The total cross sections without FSI are shown
by the dashed line. The solid line is obtained by
numerically integrating the angular distribution with FSI included.
The dash dotted curve indicates the total cross section calculated
with FSI included but assuming the angular distribution to be isotropic
(i.e. $\sigma_{tot} = 4 \pi (d\sigma/d\Omega)_{\theta_{\eta}=180^0}$).
The data of Ref. \cite{mayer} seems to indicate
a negligible forward-backward asymmetry
(which is consistent with zero within 5\% at all energies from
0 to 11 MeV above threshold) in the $p\, d \rightarrow \, ^3$He$\, \eta$
reaction. Within the two-step FSI model of the present
work, however, we find small deviations from isotropy at energies
away from threshold. Since the total cross section assuming isotropic
angular
distribution gives better agreement with data, it seems that the two step
model of the present work underestimates the cross section at forward
angles.
The data on the squared amplitude $|\,f\,|^2$ at 180$^0$ is however very
well
produced as seen earlier in Fig. 2.

\begin{figure}[h] 
\centerline{\vbox{
\psfig{file=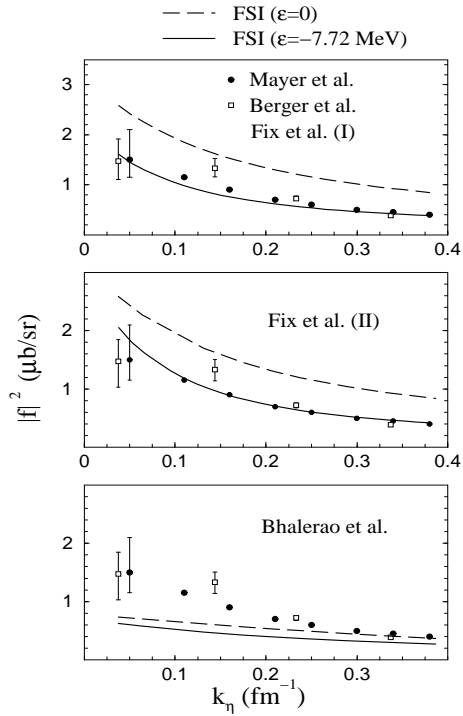,height=9.5cm,width=6cm}}}
\caption{Effects of including the binding energy of $^3$He in the
calculation
of FSI between $^3$He and $\eta$. The solid and dashed curves are
calculations including FSI with the binding energy $\varepsilon$ =
-7.718 and zero respectively (see Eq.(\protect\ref{tfsi})). The effect
is shown for three different elementary t-matrices, 
$t_{\eta \, N\, \rightarrow \, \eta \, N}$.}
\end{figure}
 
Finally, in Fig. 6 we study the effects on the
$p\, d \rightarrow \, ^3$He$\, \eta$
reaction, of including the FSI which
incorporates the binding effects in $^3$He. Once again we plot the
amplitude squared $|f|^2$ calculated for three different elementary
t-matrices, $t_{\eta \, N\, \rightarrow \, \eta \, N}$. 
The solid curves are the calculations using
$\varepsilon$ = -7.718 MeV in Eq. (\ref{tfsi}) for $T_{\eta\,^3He}$. The
dashed curves represent the calculation with $\varepsilon = 0$, i.e.
no binding effects in the final state interaction. The
calculated results without
the binding effects are larger than those which include them. The
magnitude of the increase, however, depends upon the elementary t-matrix,
$t_{\eta \, N \, \rightarrow \, \eta \, N}$, under consideration.
It is nearly a factor of 2 when we use  the t-matrix of Ref. \cite{fix},
with parameter set (I), a factor of 1.25 when we use the same t-matrix with parameter
set (II)
and negligibly small for the t-matrix of Ref. \cite {bhal}.

\section{Summary}
The $p\, d \rightarrow \, ^3$He$\, \eta$
reaction has been studied in the present
work within a two step model, incorporating the final state interaction
(FSI)
of the $^3$He nucleus and $\eta$ meson in a rigorous way. The peculiar
behaviour of the cross section for
this $\eta$ producing reaction as compared to a similar
pion production reaction $p\, d \rightarrow \,^3$He$\,\pi^0$ is seen to
originate due to the interaction between $^3$He and $\eta$. The
FSI changes the energy dependence of the squared amplitude
from a constant (without FSI) to one which
falls by a factor of about 4, from threshold to 11 MeV above threshold.
We incorporate the FSI through an off-shell T-matrix for $\eta\,^3$He
elastic
scattering. This T-matrix is evaluated by numerically solving few body
equations which include the nuclear binding effects. Both the off-shell
as well binding effects in $\eta\,^3$He scattering are found to be
important in the calculation of the $p\, d \rightarrow \, ^3$He$\, \eta$
reaction near threshold. Earlier investigations of this
reaction involving on-shell and approximate ways of calculating the
FSI should hence be treated with caution.

The $\eta$ nucleus interaction has generated a lot of interest in the
past few years,
particularly due to the possibility of forming $\eta$ mesic nuclei.
Since it is difficult to obtain data on elementary $\eta$ nucleon
scattering, little is known about the $\eta N$ interaction. The
scattering length in $\eta N$ scattering is a much debated quantity
and different estimates and limiting values (for the
possible formation of an $\eta$ mesic nucleus) of this parameter
exist in literature. Within the models used in  the present work we
find that values of the $\eta N$ scattering length, Re $a_{\eta\,N} \, \sim$ 0.75 to 0.9
and Im $a_{\eta\,N} \, \sim$ 0.3 to 0.4
lead to a good reproduction of the $p\, d \rightarrow \, ^3$He$\, \eta$
data near threshold. 
 
\vskip1cm
\noindent
{\Large \bf Acknowledgements}
\vskip0.3cm
 
The authors wish to thank Prof. R. A. Arndt for providing the code
required for
the calculation of the $ p\, p\, \rightarrow \, \pi^+  \, d$ amplitudes
and
Profs. R. S. Bhalerao and E. Oset for useful
discussions. One of the authors (K.P.K.) gratefully acknowledges the
award of a research fellowship from the Department of Atomic Energy of
the Government of India.

\vskip1cm
\noindent
\setcounter{equation}{0}
\renewcommand{\theequation}{A.\arabic{equation}} 
{\Large \bf Appendix}
\vskip0.5cm

In what follows, we discuss in  detail the constituents of the Born
amplitude
(Eq. (\ref{ampli})) for the $p\, d \rightarrow \, ^3$He$\, \eta$
reaction.
To start with, we write the deuteron wave function in Eq. (\ref{ampli})
as,
\begin{eqnarray}
&&\sum  <\, p n\,|\,d\,> \, = \,{1 \over \surd 2} \, \Bigr \{ \,
\sum_{m_n} <\, 1/2\,\, m_p\,\, 1/2\,\, m_n \,\, | \, 1\,\, m_2\,>
\, {\phi_0^d (p_1) \over \surd 4\pi } 
\\ \nonumber
&& + \sum_{m_l} <1/2\, \, m_p\, 1/2\, \, m_n \, | \, 1\, M_s>
<1 \,\, M_s\,\, 2 \,\, m_l\,|\, 1\,\, m_2>Y_{2,m_l}(\hat{p_1})\, \,
 \phi_2^d (p_1) \,\Bigr \}
\end {eqnarray}
where $m_n$ and  $m_p$ are the spin projections of the off-shell neutron
and
proton respectively in the intermediate state and $m_2$ is the spin
projection of
the target deuteron. The factor ${ 1 \over \surd 2}$ comes from isospin
overlap.
$\phi_l^d (p_1)$ is the deuteron wave function in the $l^{th}$ partial
wave and
$\vec {p_1}$ is the relative momentum of the p-n pair inside the
deuteron.
Writing
\begin{equation}
\phi_0^d (p_1) = ( 2 / \pi )^{1 \over 2} \, \, \sum_{j=1}^n \, \, {C_j
\over p_1^2 + m_j^2}
\end {equation}
\begin{equation}
\phi_2^d (p_1) = ( 2 / \pi)^{1 \over 2} \, \,  \sum_{j=1}^n \, \,
{D_j \over p_1^2 + m_j^2},
\end {equation}
the parameters $C_j$ , $D_j$ and $m_j$ for the Paris potential are given
in Ref. \cite{paris}.

The $^3$He wave function is written as,
\begin{eqnarray}
&&\sum \, <\, ^3He \, | \, p \, d\,> = \sum_{m_2^\prime , m_p^\prime}
<\, 1 \, \, m_2^\prime \, \, 1/2 \, \, m_p^\prime \, | \, 1/2 \, \, m_3
\, >
{ \chi_0(p_2) \over \surd 4 \pi } +
\\ \nonumber
&&\sum_{m_l^\prime}
<1 \, m_2^\prime \, \, 1/2 \, \, m_p^\prime \, | \,  3/2 \, \, m>
<2 \, \, m_l^\prime \, \, 3/2 \, \, m \, | \, 1/2 \, \, m_3>
\chi_2 (p_2) \, Y_{2 , m_l^\prime} (\hat {p_2})
\end{eqnarray}
where $m_2^\prime$ and $m_p^\prime$ are the spin projections of the
off-shell
deuteron and proton respectively. The spin projection of $^3$He is $m_3$
as shown in Fig. 1. $\chi_l (p_2)$ is the helium wave function in the
$l^{th}$ partial wave
and  $\vec {p_2}$ is the relative momentum of the p-d pair inside $^3$He.
\begin{equation}
\chi_l (p_2) = \, \, \sum_{i=1}^n \, \, {a_i \over p_2^2 + m_i^2}
\end {equation}
The parameters $a_i$ and $m_i$ given in Ref. \cite{helium} are chosen
corresponding
to p-d clustering in $^3$He. The normalization of the wave function is
such that,
\begin{equation}
\int p_2^2 dp_2 \, \, \{ \chi_0 (p_2)^2 + \, \, \chi_2 (p_2)^2 \} = 1.5
\end{equation} \\
 
The four momentum k$_\pi$ = \{ E$_\pi$ , $\vec {k_\pi}$ \} appearing in
the pion
propagator in Eq. (\ref{ampli}) is written using energy and momentum
conservation at the $ \pi^+ \, n \rightarrow \eta \, p$ and
$ p \, p \, \rightarrow \, \pi^+ \,d $ vertices respectively. Thus,
\begin{equation}
E_\pi = E_\eta + { 1 \over 3} E_{He} - { 1 \over 2 } E_d
\end{equation}
and
\begin{equation}
\vec {k_\pi} = { \vec {k_p} \over 2 } + { 2 \over 3} \vec {k_\eta} +
\vec{p_1} + \vec{p_2}
\end{equation}

\end{document}